\documentclass[aoas]{imsart}

\RequirePackage{amsthm,amsmath,amsfonts,amssymb,graphicx,bm}
\RequirePackage[authoryear]{natbib}
\RequirePackage{url}
\RequirePackage{color}
\RequirePackage{comment}
\newcommand{\hlc}{{\text{LC}^+}}
\startlocaldefs

\endlocaldefs

\begin{document}

\begin{frontmatter}
\title{Bayesian Region Selection and Prediction in Poisson Regression with Spatially Dependent Global-Local Shrinkage Prior}
\runtitle{Spatially-Dependent Global-Local Shrinkage}

\begin{aug}
\author[A]{\fnms{Zihan}~\snm{Zhu}\ead[label=e1]{zihan.zhu@yale.edu}},
\author[B]{\fnms{Xueying}~\snm{Tang}\ead[label=e2]{xytang@arizona.edu}}
\and
\author[C]{\fnms{Shuang}~\snm{Zhou}\ead[label=e3]{szhou98@asu.edu}}
\address[A]{Department of Biostatistics, Yale School of Public Health\printead[presep={,\ }]{e1}}

\address[B]{Department of Mathematics, University of Arizona\printead[presep={,\ }]{e2}}
\address[C]{School of Mathematical and Statistical Sciences, Arizona State University\printead[presep={,\ }]{e3}}
\end{aug}

\begin{abstract}
High-dimensional spatially correlated covariates are common in regression models encountered in environmental sciences and other fields. In such models, the regression coefficients often exhibit a sparse structure with spatial dependence. Although standard variable selection approaches can help detect the sparse structure, incorporating the dependence into variable selection helps recover spatially contiguous signals and improves prediction accuracy. Motivated by a real-world challenge in hurricane count prediction, we propose a novel neighborhood-structured global-local shrinkage prior for prediction and region selection in Poisson regression with spatial covariates. The proposed prior combines the Conditional Auto-Regressive (CAR) prior with a Super Heavy-tailed prior to introduce spatial dependence among the coefficients while ensuring appropriate shrinkage effects for covariate selection. We develop an efficient  Metropolis-within-Gibbs sampler for computation that accommodates the count data. Extensive simulation studies demonstrate that the proposed model excels when signals are weak and adjacent and the spatial dependence in covariates is strong. In the application of hurricane prediction from the north Atlantic, our method outperforms traditional regression-based approaches and rivals the benchmark ``oracle" model.
\end{abstract}


\end{frontmatter}



\section{Introduction}
In many scientific applications, spatially structured covariates arise from area-aggregated measurements of environmental or physical quantities over regular areal units with well-defined spatial boundaries \citep{banerjee2004spatial, cressie2015statistics}. These covariates are commonly employed in applications including disease incidence~\citep{wakefield2007disease}, crop yield~\citep{lobell2007historical}, and storm frequency~\citep{elsner2006prediction,vecchi2011statistical}, where the goals are to predict the outcome and to identify the influential subsets of regions~\citep{li2015spatial,meyer2019importance,zhao2023bayesian}. Recent advances in remote sensing and climate reanalysis provide these covariates at increasingly fine spatial resolutions, often yielding far more regional predictors than available samples~\citep{wulder2004high,pettorelli2018satellite,alsafadi2023high}. Motivated by high-resolution environmental predictors measured on areal lattices, we consider sparse region selection and prediction when covariates are spatially dependent and the number of regions is large relative to the sample size. We develop a Bayesian framework that jointly provides accurate prediction and data-driven identification of influential spatial regions, while explicitly leveraging neighborhood structure. 

Our motivating example focuses on the prediction of the North Atlantic Hurricane counts, where improved forecast can mitigate substantial societal and economic losses. Existing regression-based approaches, such as the Colorado State University (CSU) model\footnote{See http://hurricane.atmos.colostate.edu/ for more information.} and the University of Arizona (UA) model \citep{davis2015new}, fit Poisson regressions to hurricane counts with sea surface temperature (SST) as the primary predictor. SST data were measured using a combination of in situ sensors (e.g., buoys and ships) and satellite-based radiometers \citep{kennedy2014review,gentemann2020fluxsat}. Statistical methods such as spatio-temporal kriging \citep{montero2015spatial}, low-rank tensor completion \citep{wang2014low} and Bayesian hierarchical smoothing \citep{zheng2023empirical} are often used to reconstruct the complete SST fields. As a result, the reconstructed SSTs span over tens of thousands of regular areal units (e.g., $2^\circ \times 2^\circ$ grid cells), yielding high-dimensional region-level covariates with strong spatial dependence. In such settings, the underlying regression signals are typically sparse and spatially clustered, posing substantial challenges for variable selection and estimation. In both CSU and UA models,  the region over which SST affects hurricane count was determined based on expert knowledge and regression models were fitted using the averaged SST over selected region. This ad hoc selection limits their adaptability to other basins such as the Eastern or Central Pacific. Hence, it is essential and necessary to develop a fully data-driven framework that adaptively selects influential regions from spatially correlated covariates for hurricane prediction.

Region selection performs variable selection while also accounting for dependence structure among variables. In variable selection, coefficients are typically assumed to be conditionally independent. In contrast, region selection involves predictors measured across multiple spatial locations that form a spatial domain with known neighborhood relationships. The objective is to identify spatially contiguous regions, that is, groups of neighboring locations, whose associated measurements are related to the outcome. This setting requires addressing both sparsity, since only a small proportion of regions are expected to be truly relevant, and spatial correlation, since nonzero effects tend to occur in clusters.  Standard variable selection procedures do not address these two features jointly.

Bayesian sparsity-inducing priors for variable selection are often formulated in two groups: (1) as discrete mixture models (e.g., spike-and-slab priors), in which each coefficient is modeled with a mixture of a Dirac measure at zero and a continuous distribution allowing large values \citep{mitchell1988bayesian,george1993variable,george1997approaches, ishwaran2005spike}, and (2) as continuous global-local shrinkage priors, in which each coefficient is expressed as a normal scale mixture with a global parameter controlling overall sparsity and local parameters allowing individual coefficients to escape shrinkage \citep{polson2010shrink,carvalho2010horseshoe,bhadra2017horseshoe+}. The discrete mixture type of priors are conceptually appealing with direct variable selection but require latent binary indicators, which become computationally burdensome in high-dimensional settings \citep{o2009review,scott2010bayes}. The continuous shrinkage priors avoid discrete indicators and adaptively shrink coefficients toward zero. However, both classes typically assume independence between variables thus ignore spatial dependence, limiting their capability of variable selection when signals exhibit regional adjacency.

Spatial dependence is commonly introduced through Gaussian field and Gaussian Markov random field (GMRF) priors.  Gaussian field models encode correlation through distance-based covariance functions \citep{lindgren2011explicit}, while Gaussian Markov random fields (GMRFs) such as conditional autoregressive (CAR) priors \citep{besag1991bayesian,rue2005gaussian} capture neighborhood-based dependence for areal data. These priors promote spatial smoothness but do not induce sparsity on their own. Several recent extensions have been proposed to incorporate spatial dependence into variable selection procedures. For instance, \citet{goldsmith2014smooth} introduced a CAR–Ising prior that achieves spatially clustered selection, but may require computationally intensive updates for binary indicators. \citet{kang2018scalar} proposed a continuous selection prior based on soft-thresholded Gaussian processes, which retains spatial smoothing via a CAR prior while avoiding discrete indicators. Both methods demonstrate the value of spatial structure in selection, but they are primarily designed for moderate-resolution scalar-on-image regression.

Complementing these indicator-based or thresholding approaches, a growing literature develops dependent continuous shrinkage priors that induce structured sparsity while avoiding discrete model indicators. For spatial signals, continuous global–local priors have been adapted to borrow strength across neighboring locations, e.g., the spatial horseshoe framework of \cite{jhuang2019spatial}, which couples shrinkage behavior across sites while retaining heavy-tailed, near-zero–favoring marginals. Related work on shrinkage-prior Markov random fields places global–local shrinkage on GMRF increments to obtain locally adaptive smoothing over graphs \citep{faulkner2017locally}. More generally, structured shrinkage priors allow coefficient collections (vectors, matrices, or tensors) to exhibit dependence while preserving marginal sparsity \citep{griffin2024structured}, and Gaussian-process-driven constructions can induce dependence in the local variance components to encourage region-sparse patterns \citep{wu2019dependent}. In parallel, dependence can be introduced at the selection stage, for example by modeling spatially varying inclusion probabilities to promote coherent selection in high-dimensional GLMs \citep{leach2022incorporating}. Finally, when the dependence graph is itself uncertain, recent Bayesian frameworks can jointly select predictors and learn an underlying network structure \citep{peterson2016joint,cao2024consistent}.

In this paper, we propose a spatially dependent global-local (SGL) shrinkage prior that unifies spatial smoothing and adaptive shrinkage in a continuous framework. By integrating a CAR prior to encode spatial dependence with a global-local hierarchy to induce sparsity, our approach eliminates the need for discrete variable selection indicators, scales efficiently to thousands of spatial locations, and yields interpretable control over both marginal shrinkage and spatial information sharing. This makes the method particularly well-suited for high-dimensional count data regression with regionally adjacent effects, providing a computationally tractable and statistically principled framework for prediction and sparse region selection.

The rest of this paper is organized as follows. Section \ref{Sec:data} describes the hurricane prediction dataset that motivates our methodology. In Section \ref{Sec:model}, we introduce the formulation and underlying mechanism of the SGL prior. Simulation studies in Section~4 exhibit the empirical performance of the proposed method. Section~5 presents a real-data application to North Atlantic hurricane prediction. We conclude with a discussion in Section~6.

\section{Data Description and Hurricane Prediction}\label{Sec:data}

Tropical cyclones are among the most devastating natural hazards worldwide, causing significant loss of life, property damage, and economic disruption. In particular, hurricanes in the North Atlantic basin have profound impacts on coastal communities in the United States, the Caribbean, and Central America. According to the National Hurricane Center\footnote{\url{https://www.nhc.noaa.gov/}}, damages from individual hurricanes can exceed tens of billions of dollars, as exemplified by Hurricane Katrina in 2005 and Hurricane Harvey in 2017. Accurate and timely prediction of hurricane activity is thus a matter of both scientific urgency and public interest. 

A key quantity in hurricane activity forecast is annual hurricane counts, i.e., the number of hurricane that occur in a given year. These counts serve as important indicators of seasonal cyclone activity and are closely monitored by agencies such as the National Oceanic and Atmospheric Administration (NOAA). Forecasts of annual hurricane counts inform disaster preparedness strategies, insurance risk assessments, and emergency response planning \citep{emanuel2005increasing,mendez2023links}. As climate variability and anthropogenic forcing continue to influence ocean-atmosphere dynamics, understanding and modeling the drivers of interannual hurricane variability has become increasingly critical.

From a statistical perspective, the major challenge of hurricane count prediction lies in the fact that the underlying climatic drivers are high-dimensional, spatially structured, and often highly correlated. This necessitates the development of flexible, interpretable, and computationally tractable models that can accommodate both count-valued outcomes and complex spatial covariates. In this paper, our goal is to develop a novel scalable Bayesian model that accounts for spatial dependence in region selection, aiming to enhance prediction accuracy and identify regions with significant contributions to hurricane genesis. To operationalize this framework, we curate a dataset consisting of annual hurricane counts and climate-related predictors over the North Atlantic basin from 1950 to 2013 ($n=64$ years). The annual hurricane counts $\bm{y}$ serve as the response variable of interest, while the predictors include the Multivariate El Ni\~no--Southern Oscillation Index with the Atlantic Multidecadal Oscillation (MEI\_AMO), zonal pseudo--wind stress (PWS), and sea surface temperature (SST), following previous studies \citep{vecchi2011statistical,davis2015new,chan2021improved}. All data are obtained from the Hurricane Research Division of NOAA\footnote{\url{http://www.aoml.noaa.gov/hrd/tcfaq/E11.html}}. MEI\_AMO and PWS are scalar-valued summaries derived from established climate indices and regionally averaged quantities, respectively. These indices summarize basin-scale atmosphere–ocean conditions known to modulate hurricane variability through thermodynamic and dynamical pathways. In contrast, SST exhibits rich spatial variation,  which provides geographically resolved information on ocean heat content and surface forcing relevant to cyclone genesis and intensification.

\begin{figure}[htp!]
    \centering
    \includegraphics[width=0.8\linewidth]{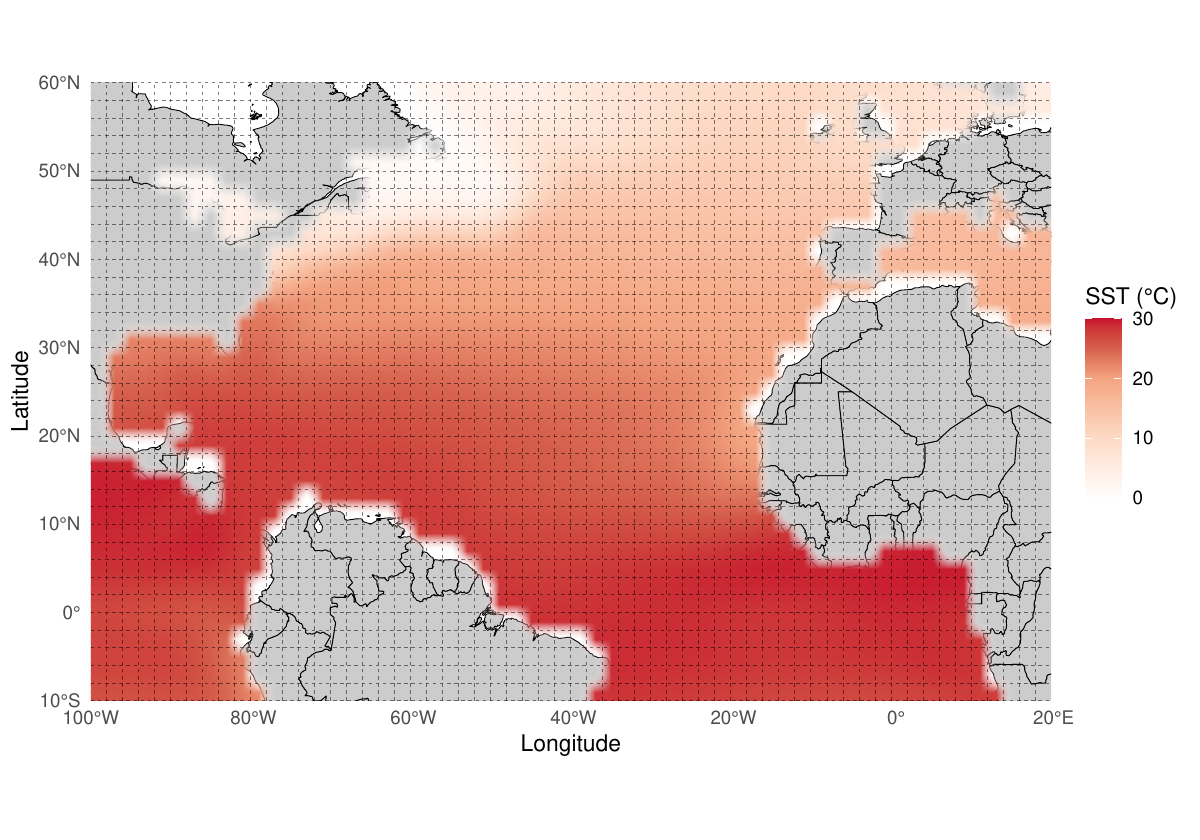}
    \caption{Average March-to-May SST in $2^{\circ} \times 2^{\circ}$ grid between 2007 and 2017.}
    \label{fig:ASST}
\end{figure}

We treat SST as a spatial covariate observed on a two-dimensional $2^\circ \times 2^\circ$ latitude-longitude grid spanning the tropical Atlantic basin. SST fields are obtained from the ERSST v3b dataset \citep{smith2008improvements}, which reconstructs historical SST values using ship- and buoy-based observations combined with empirical orthogonal function analysis and optimal interpolation. For each year, we construct a preseason SST summary by averaging monthly SST over March–May at each grid cell, yielding a spatial covariate vector $x_i = (x_{1,i},\ldots, x_{J,i})'$, where $J$ denotes the number of retained ocean grid cells after excluding land and missing values. Figure~\ref{fig:ASST} visualizes the spatial structure of SST by displaying the mean March–May SST field over a representative climatological period. Each SST grid cell represents a distinct spatial location, and the entire spatial domain is discretized into a regular lattice of $350 \times 600$ locations.  The plot highlights the strong spatial smoothness and large-scale gradients that motivate explicit dependence modeling across neighboring cells. 

Prior to model fitting, we standardize each predictor to have mean zero and unit variance across years. Standardization improves numerical stability, ensures comparability of effect sizes across heterogeneous predictors, and isolates the inferential focus on identifying a sparse subset of spatial regions whose SST variability is most predictive of annual hurricane activity.

\section{Model}\label{Sec:model}
Let $\bm{y} = (y_1, \cdots, y_n)'$ denote the vector of response counts for $n$ observations. For the $i$-th observation, let $\bm{w}_i = (w_{1,i}, \cdots, w_{K,i})'$ represent $K$ scalar covariates and $\bm{x}_i = (x_{1,i}, \cdots, x_{J,i})'$ denote a vector of spatial covariates measured across $J$ regions. The spatial structure is defined through a $J \times J$ adjacency matrix $\bm{A}$, where $A_{j,k} = 1$ if regions $j$ and $k$ are neighbors and $A_{j,k} = 0$ otherwise. The associated degree matrix $\bm{D}$ is a $J \times J$ diagonal matrix with entries $D_{j,j} = \sum_{k=1}^J A_{j,k}$. In the hurricane prediction problem, the scalar covariates are PWS and MEI\_AMO and the spatial covariates are SSTs, measured in regular grid. Two regions are neighbors if they share a common border. Each region has two to four first-order neighbors. 

We assume that $y_i$ follows a Poisson distribution with mean $\theta_i$:
\begin{equation}
    \label{eq::edm}
    f(y_i | \theta_i) = \frac{\theta_i^{y_i} e^{-\theta_i}}{y_i!},
\end{equation}
where $\theta_i$ is parametrized via a log-linear model,
\begin{equation}
    \label{eq::link}
    \log(\theta_i) = \bm{w}_i' \bm{\alpha} + \bm{x}_i' \bm{\beta}.
\end{equation}
Here, $\bm{\alpha}$ is a $K$-dimensional coefficient vector, capturing the effects of the scalar covariates $\bm w$, and $\bm{\beta}$ is a $J$-dimensional coefficient vector, representing the effects associated with the spatial covariates. We focus on settings where the spatial effects $\bm{\beta}$ are not only sparse but also exhibit spatial structure where nonzero coefficients are geographically clustered. This spatial structure imposes a dual requirement on the prior: it must enable sufficient shrinkage to eliminate noise, while preserving spatially correlated local smoothness to retain contiguous active regions.

In what follows, Section~3.1 introduces the core prior structure, decomposing each coefficient into global, local, and spatial components, with spatial dependence encoded via CAR priors. Section~3.2 discusses prior specifications for the global and local shrinkage parameters, motivates the use of log-Cauchy priors, and analyzes their effect on the induced covariance structure. Section~3.3 completes the model by specifying priors for remaining parameters and presenting the full hierarchical formulation. Section~3.4 describes a Metropolis-within-Gibbs sampler with adaptive tuning for efficient posterior computation. Section~3.5 outlines prediction and region selection based on posterior summaries of the spatial effects.

\subsection{Spatially Dependent Global-Local Shrinkage Prior}
To incorporate both sparsity and spatial dependence in $\bm{\beta}$, we decompose each $\beta_j$ as a product of three independent parameters,
\begin{equation}
\label{eq::beta_decom}
    \beta_j   = \tau \lambda_j \tilde{\beta}_j, \ \tau>0,\ \lambda_j>0
\end{equation}
where $\tilde{\beta}_j$ is used to incorporate spatial dependence, and $\tau$ as well as $\lambda_j$ are used to induce shrinkage effect and sparsity on $\beta_j$. Specifically, $\tau$ controls global shrinkage across all regions, while $\lambda_j$ enables local adaptivity to retain large signals. Prior choices for $\tau$ and $\lambda_j$ are discussed in detail in Section~3.2.

To incorporate spatial dependence, we assign $\bm{\tilde{\beta}} = (\tilde \beta_1, \cdots, \tilde \beta_J)'$ a CAR prior~\citep{besag1991bayesian,rue2005gaussian}:
\begin{equation}
    \label{eq::matrixCAR}
    \bm{\tilde \beta} \sim N_J\left(\bm{0}, (\bm{D} - \rho \bm{A})^{-1}\right),
\end{equation}
where $\rho \in [0,1)$ controls the strength of spatial correlation. Larger values of $\rho$ correspond to stronger spatial correlation. The range of $\rho$ ensures the precision matrix $(\bm{D} - \rho \bm{A})$ is positive definite and the prior is proper. The conditional form of the CAR prior reveals the spatial dependence structure more clearly:
\begin{equation}
    \label{eq::CAR}
    \tilde \beta_j \mid \bm{\tilde \beta}_{-j} \sim N \left( \frac{\rho}{D_{j,j}} \sum_{k=1}^J A_{j,k} \tilde \beta_k,\ \frac{1}{D_{j,j}} \right),
\end{equation}
where $\bm{\tilde \beta}_{-j} = (\tilde \beta_1,\cdots,\tilde \beta_{j-1}, \tilde \beta_{j+1},\cdots, \tilde \beta_J)'$. This implies that the prior mean of $\tilde \beta_j$ is the average of its neighbors’ values multiplied by $\rho$.

The structure defined in \eqref{eq::beta_decom} generalizes the classical global-local shrinkage prior \citep{polson2010shrink}, which models each $\beta_j$ as the product of a global scale $\tau$, a local scale $\lambda_j$, and an independent standard normal variable $\tilde{\beta}_j \sim N(0,1)$. The independent assumption on $\bm{\tilde{\beta}}$ induces shrinkage separately across regions. It may fail to capture the spatial structure in $\bm{\beta}$. To address this, we replace the independent Gaussian components with a spatially correlated vector $\bm{\tilde{\beta}} \sim N_J(\bm{0}, (\bm{D} - \rho \bm{A})^{-1})$, thereby coupling the shrinkage mechanism with spatial smoothing. Similar extensions have been explored in dynamic settings \citep{kowal2019dynamic}, but have not been applied in spatial region selection context.

\subsection{Priors on Global and Local Shrinkage Parameters}
In the spatially dependent global-local (SGL) prior defined in \eqref{eq::beta_decom} and \eqref{eq::matrixCAR}, the sparsity of $\bm{\beta}$ is governed by the global scale  $\tau$ and the local scales $\{\lambda_j\}$. They control not only the marginal shrinkage of each $\beta_j$, but also the extent to which spatial dependence in $\bm{\tilde{\beta}}$ is transmitted to $\bm{\beta}$. Consequently, the priors on $\tau$ and $\{\lambda_j\}$ should possess both (i) strong global shrinkage to suppress diffuse spatial noise and (ii) sufficiently heavy tails to preserve large effects in truly active regions. 

Several commonly used prior choices for global–local shrinkage address these requirements only partially. The truncated Cauchy prior \citep{van2017adaptive} \begin{equation}
    \pi(\tau) \propto \frac{1}{1+\tau^2} \cdot I\left(\frac{1}{J} \leq \tau \leq 1\right),
\end{equation} is a common choice for the global scale $\tau$, concentrating mass near a small lower bound and preventing overshrinkage. In ultra-sparse spatial settings, however, this bounded support can be restrictive, because suppressing spatially propagated noise may require more extreme global shrinkage than the truncation permits. Similarly, the half-Cauchy prior \citep{carvalho2010horseshoe}
\begin{equation}
    \pi(\lambda_j) = \frac{2}{\pi(1+\lambda_j^2)}, \quad \lambda_j > 0,
\end{equation}
is a standard choice for local scales $\lambda_j$ and performs well in classical global-local frameworks, but it may be insufficiently heavy-tailed when $\tau$ is driven very close to zero. In that regime, preserving large signals demands stronger tail robustness for $\lambda_j$ than the half-Cauchy can provide.

We adopt the log-Cauchy (LC) prior~\citep{zhu2025modeling} for both $\tau$ and each $\lambda_j$, defined as
\begin{equation}
    \label{eq::prior_on_lambda}
    \pi(x) = \frac{1}{x} \cdot \frac{1}{\pi^2 + \log^2(x)}, \quad x > 0.
\end{equation}
The LC prior has an unbounded density at the origin and super-heavy, polynomially decaying tails, placing substantial prior mass both near zero and at large magnitudes. These features are particularly well suited to spatially dependent global–local (SGL) shrinkage, where aggressive global regularization and selective local escape must be achieved simultaneously. Specifically, the global scale $\tau$ should strongly suppress diffuse spatial noise, whereas the local scales $\lambda_j$ remain heavy-tailed enough to retain large effects in active regions. The LC prior provides a coherent mechanism for balancing these competing objectives \citep{zhu2025modeling}.

Beyond marginal shrinkage, the SGL prior also induces dependence between region-specific effects $\beta_j$ and $\beta_k$ via the spatial dependence encoded in $\tilde{\beta}_j$. This covariance is further modulated by the product $\tau \lambda_j \cdot \tau \lambda_k$, effectively shrinking both marginal variances and pairwise covariances. Thus, appropriate prior choices for $\tau$ and $\lambda_j$ are essential not only for sparsity in effect sizes, but also for controlling the structure and strength of spatial dependencies. In the following, we illustrate how the joint behavior of $\tau$ and $\lambda_j$ influences the induced covariance structure under different scenarios.

To better understand the role of $\tau$ and $\lambda_j$ in modulating both marginal effects and spatial interactions, we examine the induced covariance structure under the SGL prior. In particular, we analyze the conditional covariance between region-specific effects $\beta_j$ and $\beta_k$, given fixed values of $\tau$, $\lambda_j$, $\lambda_k$, and $\rho$. For any $j, k \in \{1, \ldots, J\}$ with $j \ne k$, the conditional covariance is given by
\begin{equation}
\label{eq::cov_prior_beta}
\operatorname{Cov}(\beta_j, \beta_k \mid \tau, \lambda_j, \lambda_k, \rho) 
= \frac{\tau^2 \lambda_j \lambda_k}{\sqrt{D_{j,j} D_{k,k}}} \sum_{l=1}^{\infty} \frac{(A^l)_{j,k}}{\sqrt{D_{j,j}^{l} D_{k,k}^{l}}} \rho^l,
\end{equation}
where $(A^l)_{j,k}$ denotes the $(j,k)$-th entry of the matrix power $\bm{A}^l$, representing the number of length-$l$ paths between regions $j$ and $k$, and $D_{j,j}^l$ is the $(j,j)$-th entry of $\bm{D}^l$. A detailed derivation is provided in the Supplementary Material. Equation \eqref{eq::cov_prior_beta} shows explicitly how spatial propagation (through the $\bm{A}^l$ term and $\rho^l$ decay) and shrinkage (through $\tau^2 \lambda_j \lambda_k$) jointly determine the prior covariance structure in $\bm{\beta}$. Notably, even when two regions are strongly connected in the neighborhood graph, their induced covariance can still be attenuated if $\tau$ or the relevant local scales $\lambda_j$ are small. 

For a more detailed illustration, we consider four scenarios to demonstrate the role of $\tau$, $\lambda_j$, and $\lambda_k$ in shaping the covariance landscape and the need for super heavy-tailed priors in the SGL framework.
\begin{itemize}
    \item {\bf Case 1: Regions $j$ and $k$ are neighbors; region $j$ is active but region $k$ is inactive. }

    In this case, we aim to retain the signal at region $j$ (with large $\tau \lambda_j$), while strongly shrinking region $k$ (with small $\tau \lambda_k$). Additionally, the induced covariance between $\beta_j$ and $\beta_k$ shall be close to zero to avoid spurious borrowing from an inactive neighbor. 
    Retaining the leading terms in Equation~\eqref{eq::cov_prior_beta} yields 
\begin{equation}
\label{eq::1stordern}
    \operatorname{Cov}(\beta_j,\beta_k \mid \tau,\lambda_j,\lambda_k,\rho) 
    \approx \tau^2 \lambda_j \lambda_k \left( \frac{\rho}{D_{j,j} D_{k,k}} + \frac{(A^3)_{j,k} \rho^3}{D_{j,j}^2 D_{k,k}^2} \right) 
    \le \tau^2 \lambda_j \lambda_k \left( \frac{\rho}{6} + \frac{5\rho^3}{36} \right).
\end{equation}
Thus, suppressing the active-inactive covariance hinges on making $\tau \lambda_k$ sufficiently small, even when $\lambda_j$ remains large. This requires the need for a prior on $\tau$ that induces stronger global shrinkage than classical independent global-local priors, so that $\tau^2 \lambda_j\lambda_k$ can be small.

    \item {\bf Case 2: Regions $j$ and $k$ are neighbors and both are active.}

    In this case, we aim to preserve the signals at both regions by maintaining large values of $\tau \lambda_j$ and $\tau \lambda_k$, even when $\tau$ is small as needed to suppress noise globally. Achieving this requires $\lambda_j$ and $\lambda_k$ to take very large values. This, in turn, calls for super heavy-tailed priors on $\lambda_j$ in the SGL setting. Moreover, once $\tau \lambda_j$ and $\tau \lambda_k$ are allowed to remain large, Equation~\eqref{eq::1stordern} implies that the induced covariance between neighboring active regions, which scales with $\tau^2 \lambda_j \lambda_k$ through the CAR component, is also preserved and can be substantial. Therefore, even when $\tau$ enforces strong global shrinkage to suppress noise in inactive regions, the prior on $\lambda_j$ must retain enough mass in the tails to allow for preservation of large signals and their mutual dependence across spatially adjacent areas.

\item {\bf Case 3: Regions $j$ and $k$ are neighbors and both are inactive.}

In this setting, both $\tau \lambda_j$ and $\tau \lambda_k$ are expected to be small, resulting in strong shrinkage of $\beta_j$ and $\beta_k$ individually. Consequently, the conditional covariance between them is also naturally shrunk toward zero, as the product $\tau^2 \lambda_j \lambda_k$ becomes negligible. 

\item {\bf Case 4: Regions $j$ and $k$ are not neighbors. The length of the shortest path between them is $L \ge 2$.}

From Equation~\eqref{eq::cov_prior_beta}, we observe that $(A^l)_{j,k} = 0$ for $l < L$, and the leading nonzero term in the covariance expansion occurs at path length $L$. We can thus approximate the prior covariance as
\begin{equation}
    \label{eq::not1stordern}
    \operatorname{Cov}(\beta_j,\beta_k \mid \tau, \lambda_j, \lambda_k, \rho) 
    \approx \tau^2 \lambda_j \lambda_k \cdot \frac{\rho^L}{D_{j,j} D_{k,k}} 
    \le \tau^2 \lambda_j \lambda_k \cdot \frac{\rho^L}{6},
\end{equation}
This formulation illustrates that spatial dependence under the SGL prior decays exponentially with distance, at a rate controlled by $\rho$. When $L$ is large, $\rho^L$ becomes negligible, and the resulting covariance is effectively zero—regardless of the activity status of $j$ and $k$. When $L$ is small (e.g., $L = 2$ or $3$), the contribution of $\rho^L$ remains non-negligible. In these cases, the behavior of the induced covariance approximates that of Cases 1–3, depending on whether regions $j$ and $k$ are both active, both inactive, or one active and one inactive.
\end{itemize}

Together, these four cases show that $\tau$ and $\lambda_j$ regulate both the marginal shrinkage of individual effects and the pairwise covariance between regions through their product $\tau^2 \lambda_j \lambda_k$. Achieving both strong suppression of noise and selective preservation of local signal clusters calls for priors that concentrate mass near zero while maintaining extremely heavy tails. The log-Cauchy prior fulfills both requirements, making it well-suited for balancing global sparsity and local adaptivity in our spatially dependent region-selection setting.

\subsection{Priors on Other Parameters}
In this subsection, we specify the prior distributions for the regression coefficients for the non-spatial covariates $\bm{\alpha}$ and for the CAR dependence parameter $\rho$. Since the number of scalar covariates is relatively small compared to the number of spatial effects ($K \ll J$), the dimensionality of $\bm{\alpha}$ is modest. We therefore assign each $\alpha_k$ an independent Gaussian distribution: $\alpha_k \sim N(0, \zeta^2)$, a commonly used choice in practice. While not strictly weakly informative, this prior provides mild regularization without strongly influencing posterior inference. In all numerical experiments, we set $\zeta = 1$, after standardizing all predictors to mean zero and unit variance.

We assign a uniform prior to the spatial autoregression coefficient $\rho$ in the CAR prior defined in Equation \eqref{eq::matrixCAR}, reflecting the absence of strong prior information. Specifically, we assume
\begin{equation}
\label{eq::rho_prior}
    \rho \sim \text{Unif}(0,1),
\end{equation}
subject to two natural constraints. First, we restrict $\rho \ge 0$ to reflect the expected positive spatial correlation in sea surface temperature (SST) data, driven by the physical continuity of oceanographic processes. Second, we require the precision matrix $(\bm{D} - \rho \bm{A})$ in the CAR prior to be positive definite. While this condition does not hold for arbitrary graph structures, it is satisfied for standard grid-based spatial layouts in which the underlying graph is connected and each region has a bounded degree. In particular, for the SST data used in our application, measured over a rectangular grid with each cell having at most four neighbors, $(\bm{D} - \rho \bm{A})$ remains positive definite if $\rho \in [0,1)$; see the Supplementary Material for a formal justification in this setting.

In summary, we present the proposed hierarchical model as follows:
\begin{equation}
    \label{eq::prmodel}
    \begin{aligned}
        y_i|\theta_i & \sim \text{Poisson}(\theta_i),\ i = 1,\cdots,n, \\ 
        \log(\theta_i) & =  \bm{W}\bm{\alpha}+\bm{X}\bm{\beta},\ \ \bm{\alpha}  \sim N_K(0,\zeta^2 \bm{I}_K), \\ 
        \bm{\beta}_j &=  \tau \lambda_j \tilde{\beta_j},\ j = 1,\cdots,J, 
        \ \ \bm{\tilde{\beta}}|\rho  \sim  N_J(0,(\bm D-\rho \bm A)^{-1}), \\
        \tau & \sim \text{LC},    \lambda_j  \sim \text{LC},  \rho  \sim \text{Unif}(0,1).
    \end{aligned}
\end{equation}

\subsection{Computation}
Posterior inference is performed using a Metropolis-within-Gibbs sampler. We update the regression coefficients $\bm{\alpha}$, the local scales $\bm{\lambda} =(\lambda_1, \ldots, \lambda_J)'$, the global scale $\tau$ and and the CAR dependence parameter $\rho$. For computation, we work with the reparameterization  $\bm{\beta} = \tau \bm{\Lambda} \tilde{\bm{\beta}}$, with $\bm{\Lambda} = \text{diag}(\lambda_1,\ldots, \lambda_J)$, which separates the spatial dependence (carried by $\tilde{\bm{\beta}}$) from the global-local shrinkage (carried by $\tau$ and $\bm{\lambda}$). Because the Poisson likelihood under the log link is non-conjugate with the CAR prior and the log-Cauchy scale priors, the full conditional distributions are not available in closed form. Therefore, all updates are conducted via element-wise random walk Metropolis-Hastings steps within the Gibbs sampler.

To ensure positivity constraints on $\tau$ and $\bm{\lambda}$ while maintaining stable proposals, we apply a soft approximation to the hard positivity constraint (an indicator function $1_{x>0}(\cdot)$) based on a smooth sigmoid function $1 / (1 + \exp(-\eta x))$, where $\eta$ is chosen to balance approximation accuracy and gradient smoothness~\citep{maddison2017concrete}. This allows us to avoid boundary sampling issues while preserving the heavy-tailed behavior of the log-Cauchy prior.

We incorporate an adaptive step size tuning mechanism during the burn-in period. Specifically, proposal variances are adjusted based on the observed acceptance rate every fixed number of iterations: the step size is increased if the acceptance rate exceeds 50\%, decreased if it falls below 30\%, and held constant otherwise. This improves sampling efficiency and helps stabilize convergence.

A detailed description of the full conditional distributions and the complete sampling procedure is provided in the Supplementary Material.

\subsection{Prediction and Region Selection}
Our Bayesian framework naturally supports both outcome prediction and region-level inference through posterior summaries. For a new observation with covariates $(\bm{w}_{new}$, $\bm{x}_{new})$, posterior predictive draws are obtained from 
$$y_{new}^{(s)} | \theta_{new}^{(s)} \sim \text{Poisson}(\theta_{new}^{(s)}), \quad \theta_{new}^{(s)} = \exp\left( \bm{w}_{new}^\top \bm{\alpha}^{(s)} + \bm{x}_{new}^\top \bm{\beta}^{(s)} \right),$$
where $\{ \bm{\alpha}^{(s)}, \bm{\beta}^{(s)} \}_{s=1}^S$ are the MCMC draws from the posterior. We report the posterior predictive mean $\hat{y}_{new} = \frac{1}{S} \sum_{s=1}^S y_{new}^{(s)}$ as a point prediction, and construct the 95\% highest posterior density (HPD) credible interval to quantify prediction uncertainty. For region selection, we assess the posterior distribution of the region-specific effect $\beta_j$ for each region $j$. We designate region $j$ as active if the $95\%$ HPD credible interval for $\beta_j$ does not contain zero.

\section{Simulation Study}

\subsection{Simulation Setting}
In this section, we conduct simulations to evaluate the performance of the SGL shrinkage priors in Poisson regression with spatial covariates. Specifically, we compare the performance with that of the CAR priors and independent global-local shrinkage priors in terms of parameter estimation, prediction accuracy, and region selection accuracy.

\begin{table}[htp!]
      \caption{Models compared in simulation studies.}
      \centering
      \begin{tabular}{c c c c }
        \hline
        \hline
        \textbf{Model}  & \textbf{Prior on $\tau$} & \textbf{ Prior on  $\lambda_j$} & \textbf{Prior on $\rho$} \\
        \hline
         \textbf{DLC} & $\hlc$ & $\hlc$ & $\text{Unif}(0,1)$   \\
            \textbf{DHS} &  $C^+(0,1)$ & $C^+(0,1)$ & $\text{Unif}(0,1)$    \\
      \textbf{HS}&  $C^+(0,1)$ & $C^+(0,1)$ & $\rho \equiv 0$    \\
      \textbf{LC} &  $\hlc$ & $\hlc$ & $\rho \equiv 0$   \\
           \textbf{CAR} &      $C^+(0,1)$ &  $\lambda_j\equiv 1$ & $\text{Unif}(0,1)$  \\
        \hline
        \hline
      \end{tabular}
      \label{tab::models}
    \end{table}

All of the models under consideration can be expressed in the form of \eqref{eq::beta_decom} and \eqref{eq::matrixCAR}. The key distinctions among them lie in their prior specifications for the global shrinkage parameter $\tau$, the local shrinkage parameter $\lambda_j$ and the spatial dependence parameter $\rho$, as summarized in Table \ref{tab::models}. Notably, DLC and LC employ log-Cauchy priors for $\tau,\lambda_j$, whereas DHS and HS use half-Cauchy priors. The CAR model, in contrast, enforces a spatially structured dependency while assuming no local shrinkage.

We consider a Poisson regression model in which the response variable \( y_i \) is generated as
\[
y_i \sim \text{Poisson}\big(\exp(\bm{w}_i'\bm{\alpha}^* + \bm{x}_i'\bm{\beta}^*)\big), 
\quad i = 1,\ldots,n,
\]
where \( \bm{w}_i \in \mathbb{R}^K \) denotes scalar predictors and \( \bm{x}_i \in \mathbb{R}^J \) represents a spatially indexed predictor. The true regression coefficients satisfy \( \bm{\alpha}^* \in \mathbb{R}^K \) and \( \bm{\beta}^* \in \mathbb{R}^J \). We consider sample sizes $n = 150$ and $450$, using training/test splits of $100/50$ and $400/50$, respectively. Each observation includes $K = 2$ scalar predictors and one spatial predictor measured over a regular lattice of size \(20 \times 25\) ($J = 500$) or $15 \times 15$ ($J = 225$). The corresponding adjacency matrix \( \bm{A} \) and diagonal matrix \( \bm{D} \) are constructed accordingly. The scalar predictors \( \bm{w}_i \) are generated independently from a standard normal distribution, \( \bm{w}_i \sim N_K(\bm{0}, \bm{I}) \), with true coefficients \( \bm{\alpha}^* = (-0.25, 0.25)' \). The spatial predictor \( \bm{x}_i \) is generated from a conditional autoregressive (CAR) model,
\[
\bm{x}_i \sim N_J\big(\bm{0}, (\bm{D} - \rho_x^* \bm{A})^{-1}\big),
\]
where \( \rho_x^* \) controls the strength of spatial dependence. We consider three levels of spatial autocorrelation: no spatial dependence (\( \rho_x^* = 0 \)), moderate spatial dependence (\( \rho_x^* = 0.4 \)), and strong spatial dependence (\( \rho_x^* = 0.8 \)). The generated spatial predictors are standardized column-wise and further scaled by \( \sqrt{n} \). 

\begin{figure}[htp!]
    \centering
    \includegraphics[width=14cm]{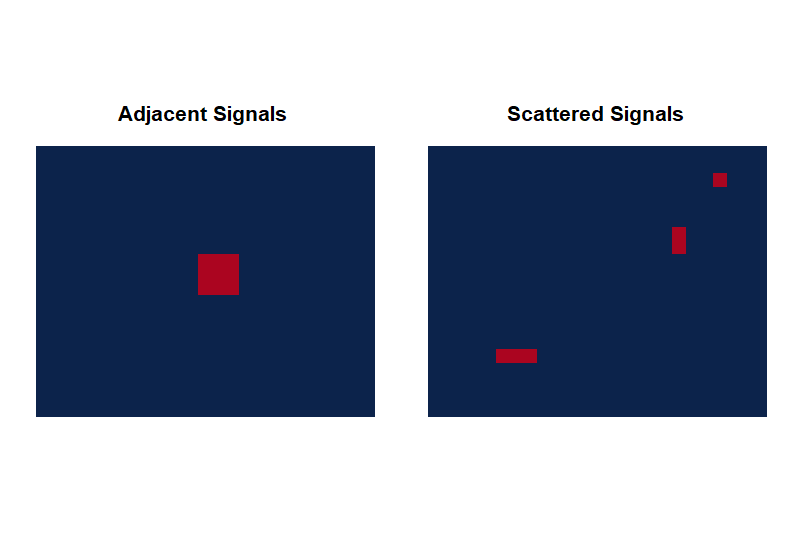}
    \caption{Patterns of active regions (red) in simulation experiments.}
    \label{fig::locations}
\end{figure}

We consider two different active region patterns for the true \( \bm{\beta}^* \), as shown in Figure \ref{fig::locations}. The adjacent signal patterns features a single contiguous signal block, while the scattered signal patterns includes several isolated nonzero regions dispersed across the spatial domain.
We expect that incorporating spatial dependence into the prior of \( \bm{\beta} \) significantly enhances estimation and selection when signals are spatially adjacent, while the benefit of incorporating dependence diminishes when signals are spatially dispersed. The true effects of \( \bm{x}_i \) are defined as:
\[
\beta_j^* = \begin{cases}
b^*,\ j \in S^*, \\ 
0,\ j \notin S^*,
\end{cases}
\]
where \( S^* \) denotes the set of indices of active regions, and \( b^* \) controls the signal strength. We consider two signal strengths: a weak signal setting with \( b^* = 6 \) and a strong signal setting with \( b^* = 8 \).

For each simulation setting \( (S^*, b^*, \rho_x^*, n, J) \), we run 100 replicates to compare the performance of DLC, LC, DHS, HS, and CAR. In each replicate, the sampling algorithm for DLC is implemented as described in Section~3.4, while the other models require only minor modifications and are therefore not discussed in detail. The MCMC algorithm for each model runs for 150{,}000 iterations with the first 60{,}000 discarded as burn-in. The acceptance rate across all models ranges between 0.25 and 0.55. To assess convergence, we initialize the chains from five distinct starting values and compute the Gelman–Rubin statistic $\hat{R}$. We find $\hat{R} < 1.1$ for all monitored parameters, indicating satisfactory mixing and convergence.

The evaluation criteria focus on three aspects: estimation accuracy, prediction accuracy, and variable selection performance. To assess the accuracy of \(\bm{\beta}\) estimation, we compute the root mean squared error (RMSE) \(\text{RMSE}_\beta = \sqrt{\frac{1}{J} (\hat{\bm{\beta}} - \bm{\beta}^*)'(\hat{\bm{\beta}} - \bm{\beta}^*)},\) where \(\hat{\bm{\beta}}\) represents the posterior mean estimator of \(\bm{\beta}\). To further evaluate the model performance in recovering true signals while shrinking noise, we report the RMSE for active and inactive regions separately as \(\text{Signal RMSE} = \sqrt{\frac{1}{|S^*|} \sum_{j \in S^*} (\hat{\beta}_j - \beta^*_j)^2}\) and \(\text{Noise RMSE} = \sqrt{\frac{1}{J - |S^*|} \sum_{j \notin S^*} (\hat{\beta}_j - \beta^*_j)^2}.\) To evaluate predictive performance, we compute the predictive root mean squared error (RMSE) \(\text{Pred\ RMSE} = \sqrt{\frac{1}{50} \sum_{i=101}^{150} (\hat{y}_i - y_i)^2},\) where \(\hat{y}_i\) denotes the expectation of the posterior predictive distribution of \( y_i \). We evaluate selection performance using the false positive rate (FPR) and false negative rate (FNR). FPR is defined as the proportion of false positives among all true negatives, while FNR is the proportion of false negatives among all true positives. 

We present results for the two signal patterns under $(n = 150, J = 500)$, while results for $(n = 150, J = 225)$ and $(n = 450, J = 225)$ are provided in the Supplementary Material.

\subsection{Results for Adjacent Signal Setting}

Table~\ref{tab::sim3lo1} summarizes the performance of all five models in terms of estimation accuracy (Signal RMSE, Noise RMSE, $\text{RMSE}_{\beta}$), predictive performance (Pred RMSE), and region selection quality (FPR and FNR) for the adjacent signal pattern.

Among the five models, CAR differs fundamentally from the rest. While it accounts for spatial structure via conditional autoregressive smoothing, it lacks a mechanism to adequately suppress noise and induce sparsity. As shown in Table~\ref{tab::sim3lo1}, CAR shows consistently inferior performance across the criteria we considered. This reinforces the need for shrinkage-based modeling when performing region selection. In subsequent discussions, we focus on other four models (DLC, DHS, LC, HS). Across all RMSEs, two consistent patterns stand out. First, spatially dependent shrinkage models (DLC and DHS) consistently outperform their independent counterparts (LC and HS), with the advantage becoming more pronounced when the covariates exhibit stronger spatial correlation. Second, prior family matters: log-Cauchy priors (DLC, LC) tend to excel under weak signals due to their sharper global shrinkage, whereas half-Cauchy priors (DHS, HS) become more favorable under strong signals, offering greater stability and reduced bias near the origin.

\begin{table}[htp!]
\centering
\resizebox{\textwidth}{!}{
\begin{tabular}{ccc|rrrrrr}
  \hline
   \hline
Signal & Corr & Model & Signal.RMSE & Noise.RMSE & Beta.RMSE & Pred.RMSE & FPR & FNR \\ 
  \hline
Weak & None & DLC & \textbf{0.859 (0.059)} & 0.065 (0.006) & \textbf{0.134 (0.010)} & \textbf{0.029 (0.003)} & \textbf{0.000} & \textbf{0.019} \\
Weak & None & DHS & 1.159 (0.055) & \textbf{0.055 (0.002)} & 0.166 (0.008) & 0.034 (0.003) & 0.020 & 0.031 \\
Weak & None & LC & 1.121 (0.106) & 0.087 (0.010) & 0.176 (0.017) & 0.036 (0.005) & 0.001 & 0.057 \\
Weak & None & HS & 1.577 (0.105) & 0.071 (0.007) & 0.224 (0.016) & 0.050 (0.007) & 0.000 & 0.238 \\
Weak & None & CAR & 4.328 (0.033) & 0.329 (0.004) & 0.668 (0.003) & 0.126 (0.010) & 0.002 & 0.078 \\

  \hline
Weak & Moderate & DLC & \textbf{0.666 (0.033)} & 0.046 (0.003) & \textbf{0.102 (0.005)} & \textbf{0.025 (0.003)} & \textbf{0.000} & \textbf{0.003} \\ 
Weak & Moderate & DHS & 0.855 (0.047) & \textbf{0.045 (0.002)} & 0.124 (0.006) & 0.030 (0.003) & 0.126 & 0.008 \\ 
Weak & Moderate & LC & 0.872 (0.079) & 0.063 (0.006) & 0.134 (0.012) & 0.028 (0.003) & \textbf{0.000} & 0.027 \\ 
Weak & Moderate & HS & 0.973 (0.065) & 0.039 (0.002) & 0.137 (0.009) & 0.029 (0.003) & \textbf{0.000} & 0.083 \\ 
Weak & Moderate & CAR & 3.872 (0.078) & 0.339 (0.007) & 0.621 (0.012) & 0.149 (0.012) & 0.006 & 0.014 \\ 
  
  \hline
  Weak & Strong & DLC & \textbf{0.670 (0.052)} & 0.047 (0.005) & \textbf{0.103 (0.008)} & \textbf{0.023 (0.003)} & 0.020 & 0.002 \\ 
Weak & Strong & DHS & 0.783 (0.059) & \textbf{0.043 (0.004)} & 0.114 (0.009) & 0.026 (0.003) & 0.505 & \textbf{0.001} \\ 
Weak & Strong & LC & 1.271 (0.127) & 0.111 (0.015) & 0.207 (0.022) & 0.043 (0.006) & \textbf{0.001} & 0.074 \\ 
Weak & Strong & HS & 1.862 (0.153) & 0.130 (0.015) & 0.285 (0.025) & 0.062 (0.008) & 0.002 & 0.182 \\ 
Weak & Strong & CAR & 3.607 (0.029) & 0.369 (0.003) & 0.608 (0.002) & 0.162 (0.012) & 0.016 & 0.003 \\ 
  \hline
  Strong & None & DLC & \textbf{0.597 (0.057)} & 0.053 (0.006) & \textbf{0.098 (0.010)} & 0.025 (0.002) & 0.001 & 0.008 \\ 
Strong & None & DHS & 0.680 (0.083) & \textbf{0.051 (0.009)} & 0.106 (0.014) & 0.024 (0.002) & 0.326 & 0.010 \\ 
Strong & None & LC & 0.675 (0.074) & 0.067 (0.006) & 0.114 (0.011) & \textbf{0.023 (0.002)} & \textbf{0.000} & 0.009 \\ 
Strong & None & HS & 1.206 (0.154) & 0.092 (0.016) & 0.189 (0.026) & 0.048 (0.009) & 0.001 & 0.083 \\ 
Strong & None & CAR & 5.467 (0.039) & 0.478 (0.006) & 0.876 (0.003) & 0.258 (0.017) & 0.022 & \textbf{0.001} \\ 
  \hline
  Strong & Moderate & DLC & 0.761 (0.111) & 0.081 (0.018) & 0.133 (0.023) & 0.038 (0.010) & 0.062 & 0.006 \\ 
Strong & Moderate & DHS & \textbf{0.546 (0.028)} & \textbf{0.040 (0.003)} & \textbf{0.085 (0.005)} & \textbf{0.022 (0.003)} & 0.740 & \textbf{0.000} \\ 
Strong & Moderate & LC & 1.027 (0.135) & 0.117 (0.020) & 0.183 (0.027) & 0.055 (0.014) & \textbf{0.001} & 0.043 \\ 
Strong & Moderate & HS & 2.807 (0.240) & 0.296 (0.032) & 0.483 (0.044) & 0.116 (0.014) & 0.007 & 0.229 \\ 
Strong & Moderate & CAR & 5.095 (0.035) & 0.503 (0.005) & 0.848 (0.003) & 0.256 (0.018) & 0.030 & \textbf{0.000} \\ 
  \hline
  Strong & Strong & DLC & 1.881 (0.232) & 0.327 (0.056) & 0.421 (0.063) & 0.116 (0.026) & 0.189 & 0.026 \\ 
Strong & Strong & DHS & \textbf{1.165 (0.177)} & \textbf{0.166 (0.040)} & \textbf{0.236 (0.045)} & \textbf{0.075 (0.024)} & 0.872 & \textbf{0.000} \\ 
Strong & Strong & LC & 2.919 (0.237) & 0.474 (0.058) & 0.628 (0.064) & 0.183 (0.033) & \textbf{0.009} & 0.227 \\ 
Strong & Strong & HS & 4.569 (0.242) & 0.664 (0.053) & 0.915 (0.060) & 0.219 (0.024) & 0.022 & 0.351 \\ 
Strong & Strong & CAR & 4.461 (0.105) & 0.498 (0.012) & 0.779 (0.017) & 0.300 (0.021) & 0.041 & \textbf{0.000} \\ 

  \hline
  \hline
\end{tabular}
}
\caption{Performance of five models under adjacent signal setting. The values in each cell are the average of 100 replicates and the values in parentheses are corresponding standard deviation of 100 replicates.}
\label{tab::sim3lo1}
\end{table}

For the estimation of regression coefficients and predictive accuracy, the patterns are closely aligned. When signals are weak, spatially dependent priors (DLC, DHS) outperform their independent counterparts (LC, HS), with DLC achieving the most accurate coefficient estimates and the lowest prediction errors across all spatial correlation levels. This advantage becomes more pronounced as spatial correlation increases, reflecting the benefit of information sharing across spatially structured effects. In contrast, independent priors are unable to adapt to the spatial configuration of signals, resulting in elevated estimation and prediction errors. When signals are strong and the covariates exhibit no spatial dependence, the independent model LC slightly outperforms all others, suggesting that incorporating spatial dependence can introduce unnecessary complexity when the underlying structure is absent. As spatial correlation increases, DLC and DHS regain their advantage, with DHS achieving the lowest beta RMSE and prediction error under strong signals and moderate to high spatial correlation. This shift likely reflects the improved stability of the half-Cauchy prior in strong-signal settings, where the singular behavior of the log-Cauchy prior near the origin can introduce estimation bias.

Signal estimation results further support these trends. In weak signal scenarios, incorporating spatial dependence allows information to be effectively shared across neighboring regions, reducing estimation error, whereas models lacking spatial structure, such as LC and HS, fail to adapt and exhibit substantially higher errors. Among spatial priors, DLC typically yields the lowest signal RMSE across different correlation levels, suggesting that the log-Cauchy prior offers more stable shrinkage in sparse regimes. As signal strength increases, the performance gap narrows, but spatial priors continue to maintain an edge, highlighting that even under strong signal conditions, modeling spatial structure remains beneficial for accurate signal recovery. For noise estimation, DHS achieves the most accurate and stable performance across all settings. It consistently yields the lowest noise RMSE, indicating its strong ability to suppress irrelevant variation. Spatial dependence plays a critical role: for both prior types, spatially dependent models (DHS and DLC) outperform their independent counterparts (HS and LC), demonstrating that incorporating spatial structure improves noise estimation. Between prior families, half-Cauchy priors (DHS, HS) generally lead to stronger noise shrinkage than log-Cauchy priors (DLC, LC), likely due to their more aggressive penalization of small coefficients. These patterns highlight that both spatial adaptivity and prior choice contribute to effective noise control.

To examine the region selection performance of the models, we take into account their performance in terms of FNR and FPR jointly.  A desirable method should achieve low FNR while maintaining tight control of FPR, particularly in high-dimensional settings where false discoveries are costly. Across all simulation settings, the DLC model consistently achieves this balance: its FPR remains low in every scenario, while its FNR is among the lowest, especially under strong spatial correlation, indicating that the model not only avoids excessive false detections but also retains strong sensitivity to weak signals. In contrast, DHS often achieves extremely low FNR—even outperforming DLC in several settings—but at the cost of substantially inflated FPR, sometimes exceeding 0.5. This pattern suggests that the half-Cauchy prior, while adaptive to strong signals, may lack sufficient regularization to suppress noise-induced false positives. Independent models (LC and HS) generally display either overly conservative or under-sensitive behavior. HS maintains nearly zero FPR across all settings, but this comes at the expense of high FNR, indicating severe under-selection. LC, while better than HS in recovering signals, still falls short of DLC in balancing specificity and sensitivity, highlighting the advantage of incorporating spatial dependence. These results underscore that effective region selection requires not only sparsity-inducing priors but also mechanisms to borrow strength across spatially related coefficients. The proposed DLC model accomplishes both, leading to superior variable selection performance under both weak and strong signal regimes.

\begin{figure}[htp!]
\centering
\includegraphics[width=0.8\textwidth]{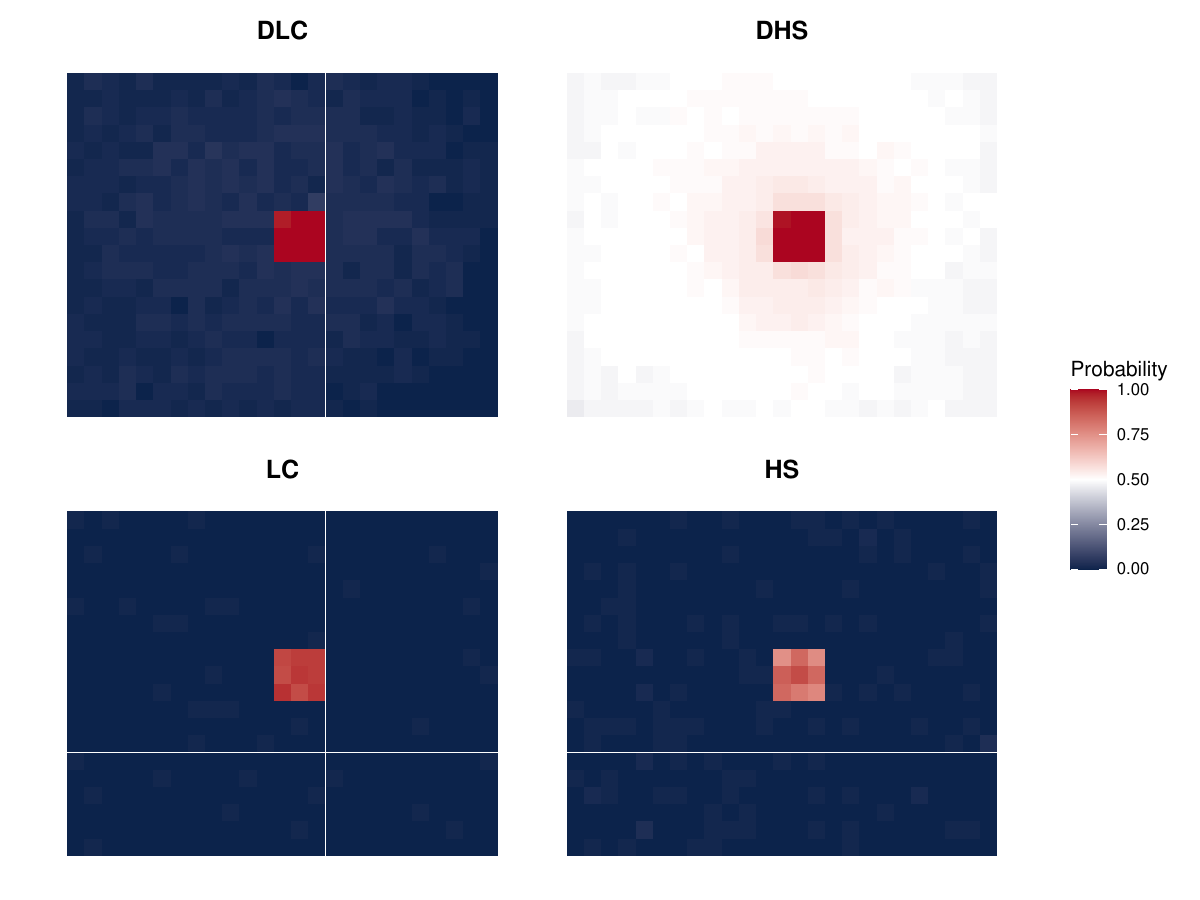} 
\caption{Heatmap of the probabilities of a region being selected as an active region under the adjacent signals setting, weak and the spatial correlation in spatially dependent predictor is moderate.}
\label{fig::po_CI_selection_lo1_b6_rho0.8}
\end{figure}

To further investigate the selection behaviors across different priors, we visualize in Figure~\ref{fig::po_CI_selection_lo1_b6_rho0.8} the selection frequency for each region across 100 replicates under weak signals and moderate spatial correlation setting. Both DLC and LC exhibit tightly concentrated selection around the true signal region, reflecting the stronger shrinkage induced by the log-Cauchy prior, which effectively suppresses noise outside the active region. In contrast, DHS shows substantial spatial diffusion, with many neighboring non-signal regions being selected at nontrivial probabilities. This explains the higher FPR observed earlier for DHS. The underlying reason is that the half-Cauchy prior in DHS, though spatially adaptive, imposes weaker shrinkage near the origin, allowing spatially correlated weak signals to propagate into surrounding noise regions. These small propagated effects often yield posterior credible intervals that exclude zero, thus triggering false positives under the CI rule. HS displays slightly more concentrated selection than DHS but still with some dispersion, consistent with its limited ability to separate weak signals from noise. 

\subsection{Results for Scattered Signal Setting}
The results of the scattered signal setting are summarized in Table~\ref{tab::sim3lo2}. As in the adjacent case, we organize the analysis around three core aspects: estimation accuracy, prediction performance, and region selection. The CAR model is included as a reference but exhibits consistently inferior performance across all metrics due to its lack of local shrinkage. Consequently, it is excluded from the detailed discussion below.

\begin{table}[htp!]
\centering
\resizebox{\textwidth}{!}{\begin{tabular}{ccc|rrrrrr}
    \hline
    \hline
Signal & Corr & Model & Signal.RMSE & Noise.RMSE & Beta.RMSE & Pred.RMSE & FPR & FNR \\ 
  \hline
Weak & None & DLC & 1.083 (0.116) & 0.087 (0.012) & 0.150 (0.017) & 0.032 (0.003) & 0.003 & 0.060 \\ 
Weak & None & DHS & 0.967 (0.052) & \textbf{0.040 (0.002)} & 0.114 (0.006) & 0.028 (0.002) & 0.010 & 0.052 \\ 
Weak & None & LC & \textbf{0.777 (0.049)} & 0.057 (0.004) & \textbf{0.104 (0.006)} & \textbf{0.026 (0.002)} & \textbf{0.000} & \textbf{0.010} \\ 
Weak & None & HS & 0.986 (0.058) & \textbf{0.040 (0.002)} & 0.116 (0.007) & 0.028 (0.002) & \textbf{0.000} & 0.078 \\ 
Weak & None & CAR & 4.998 (0.108) & 0.223 (0.006) & 0.592 (0.013) & 0.127 (0.010) & 0.000 & 0.781 \\ 
  \hline
Weak & Moderate & DLC & 0.963 (0.078) & 0.072 (0.005) & 0.130 (0.010) & 0.027 (0.003) & 0.001 & 0.043 \\ 
Weak & Moderate & DHS & 0.885 (0.041) & 0.037 (0.002) & 0.104 (0.005) & 0.023 (0.002) & 0.000 & 0.038 \\ 
Weak & Moderate & LC & \textbf{0.703 (0.029)} & 0.054 (0.003) & \textbf{0.098 (0.004)} & \textbf{0.021 (0.002)} & \textbf{0.000} & \textbf{0.003} \\ 
Weak & Moderate & HS & 0.892 (0.044) & \textbf{0.036 (0.002)} & 0.105 (0.005) & 0.024 (0.002) & \textbf{0.000} & 0.048 \\ 
Weak & Moderate & CAR & 4.813 (0.119) & 0.237 (0.007) & 0.579 (0.014) & 0.119 (0.009) & 0.000 & 0.589 \\ 
  \hline
Weak & Strong & DLC & 0.804 (0.046) & 0.047 (0.003) & \textbf{0.102 (0.006)} & 0.028 (0.003) & 0.000 & 0.010 \\ 
Weak & Strong & DHS & 0.927 (0.048) & \textbf{0.034 (0.002)} & 0.108 (0.005) & 0.026 (0.002) & \textbf{0.000} & 0.037 \\ 
Weak & Strong & LC & \textbf{0.767 (0.043)} & 0.053 (0.004) & \textbf{0.102 (0.006)} & \textbf{0.025 (0.002)} & \textbf{0.000} & \textbf{0.008} \\ 
Weak & Strong & HS & 0.935 (0.051) & \textbf{0.034 (0.002)} & 0.109 (0.006) & 0.027 (0.003) & \textbf{0.000} & 0.058 \\ 
Weak & Strong & CAR & 4.017 (0.223) & 0.203 (0.012) & 0.485 (0.027) & 0.085 (0.011) & 0.001 & 0.379 \\ 
  \hline
Strong & None & DLC & 0.876 (0.125) & 0.074 (0.013) & 0.124 (0.019) & 0.027 (0.006) & 0.008 & 0.030 \\ 
Strong & None & DHS & 0.671 (0.025) & 0.031 (0.001) & 0.081 (0.003) & \textbf{0.022 (0.002)} & 0.082 & \textbf{0.000} \\ 
Strong & None & LC & \textbf{0.598 (0.022)} & 0.045 (0.003) & 0.082 (0.003) & \textbf{0.022 (0.002)} & \textbf{0.000} & \textbf{0.000} \\ 
Strong & None & HS & 0.678 (0.024) & \textbf{0.027 (0.001)} & \textbf{0.080 (0.003)} & 0.023 (0.002) & \textbf{0.000} & \textbf{0.000} \\ 
Strong & None & CAR & 3.823 (0.472) & 0.217 (0.028) & 0.473 (0.058) & 0.128 (0.023) & 0.003 & 0.173 \\ 
  \hline
Strong & Moderate & DLC & 0.604 (0.053) & 0.050 (0.006) & 0.085 (0.008) & 0.022 (0.003) & 0.000 & 0.008 \\ 
Strong & Moderate & DHS & 0.696 (0.086) & 0.040 (0.008) & 0.088 (0.012) & \textbf{0.019 (0.002)} & 0.176 & 0.010 \\ 
Strong & Moderate & LC & \textbf{0.551 (0.021)} & 0.049 (0.002) & 0.080 (0.003) & \textbf{0.019 (0.002)} & \textbf{0.000} & \textbf{0.000} \\ 
Strong & Moderate & HS & 0.627 (0.032) & \textbf{0.028 (0.002)} & \textbf{0.075 (0.004)} & \textbf{0.019 (0.002)} & \textbf{0.000} & 0.012 \\ 
Strong & Moderate & CAR & 4.817 (0.344) & 0.279 (0.020) & 0.597 (0.043) & 0.150 (0.016) & 0.004 & 0.186 \\ 
  \hline
Strong & Strong & DLC & \textbf{0.552 (0.021)} & 0.035 (0.002) & \textbf{0.071 (0.003)} & \textbf{0.019 (0.001)} & \textbf{0.000} & \textbf{0.000} \\ 
Strong & Strong & DHS & 0.602 (0.021) & \textbf{0.028 (0.001)} & 0.073 (0.002) & \textbf{0.019 (0.001)} & 0.158 & \textbf{0.000} \\ 
Strong & Strong & LC & 0.576 (0.019) & 0.048 (0.003) & 0.082 (0.003) & 0.021 (0.002) & \textbf{0.000} & \textbf{0.000} \\ 
Strong & Strong & HS & 0.719 (0.069) & 0.034 (0.005) & 0.087 (0.009) & 0.020 (0.002) & \textbf{0.000} & 0.013 \\ 
Strong & Strong & CAR & 6.558 (0.022) & 0.373 (0.004) & 0.809 (0.001) & 0.210 (0.015) & 0.010 & 0.163 \\ 
  \hline
  \hline
\end{tabular}
}
\caption{Performance of five models under the scattered signal setting. The values in each cell represent the average over 100 replicates, with the values in parentheses indicating the corresponding standard deviation.}
\label{tab::sim3lo2}
\end{table}

Across all criteria under the scattered signal setting, independent priors (LC, HS) generally outperform their spatially dependent counterparts (DLC, DHS), especially when spatial correlation among predictors is low or moderate. This advantage is most pronounced under weak signals, where spatial shrinkage tends to oversmooth and blur the distinction between signals and noise, leading to inflated estimation bias. Prior family effects are also evident: LC maintains stability and low error across signal regimes, while HS exhibits competitive performance in coefficient estimation but can be less stable in terms of selection performance. Spatially dependent priors regain an advantage only when both signal strength and spatial correlation are high, where the imposed structure aligns with the data-generating process and supports effective borrowing of information across neighboring predictors.

For both coefficient estimation and prediction accuracy, these patterns are consistent. Under weak signals, LC achieves the lowest beta RMSE and prediction RMSE across all spatial correlation levels, followed closely by HS, confirming that in the absence of spatially structured signals, imposing spatial dependence introduces unnecessary complexity without predictive benefit. This remains true for strong signals with low or moderate spatial correlation, where LC and HS outperform spatially dependent models. In the strong-signal, strong-correlation regime, DLC and DHS match or slightly improve upon LC and HS, reflecting the benefit of spatial regularization when the correlation structure is coherent and signals are strong enough to be reliably estimated.

Signal and noise estimation results further explain these trends. In weak-signal settings, independent priors yield lower signal RMSE than spatially dependent ones, regardless of spatial correlation, underscoring the misalignment between spatial shrinkage and scattered signal configurations. Among independent priors, LC typically achieves the smallest signal RMSE, while HS performs similarly but with slightly higher variability. Noise estimation favors HS, which often attains the lowest noise RMSE, particularly under strong signals, suggesting more aggressive suppression of irrelevant variation. In contrast, spatially dependent models tend to produce larger noise RMSE in scattered settings unless signal strength and spatial correlation are both high.

Selection performance, measured by false positive rate (FPR) and false negative rate (FNR) under the credible interval criterion, also favors independent priors. Under weak signals, LC achieves near-zero FPR and FNR below 0.01 across all correlation levels, offering the most favorable trade-off. DHS and HS also control FPR effectively, but their FNRs are moderately higher, indicating reduced sensitivity to scattered weak signals. As signals strengthen, FNRs decline across models, but differences in FPR control widen: DHS exhibits elevated FPR in high-correlation, strong-signal scenarios, consistent with overselection from heightened adaptivity. LC and HS maintain near-zero FPR while retaining strong detection power, underscoring their reliability. Overall, scattered-signal regimes reward stable, non-spatial shrinkage—particularly LC—for delivering parsimonious inclusion without loss of sensitivity.

\section{Real Data Analysis}

We apply the proposed spatially dependent global-local shrinkage model to predict the annual number of named storms in the North Atlantic from 1950 to 2013 using three sets of predictors: PWS and MEI\_AMO, and spatial covariates SSTs. In addition to improving predictive accuracy, our framework aims to identify regions with the most substantial contributions to storm activity, thereby performing simultaneous prediction and region selection. Section~5.1 outlines the prediction framework and evaluation metrics. Results are presented in Section~5.2.

\subsection{Analysis Settings}

To evaluate the predictive performance of different models, we employ a recursive one-year-ahead forecasting approach \citep{west2005forecast} over the period 2001–2013. Specifically, when predicting the number of hurricanes in year \( t \) (\( 2001 \leq t \leq 2013 \)), we fit the model using all available data up to year \( t-1 \).  

In addition to the four shrinkage models evaluated in the simulation studies (DLC, LC, DHS, HS), we benchmark our approach against three established methods for hurricane prediction: a purely statistical model, Poisson regression with elastic net penalty (EN); a climatology baseline, the five-year moving average (MA); and a physically informed model—the University of Arizona (UA) model~\citep{davis2015new}. The EN model extends the Poisson regression framework commonly used in hurricane prediction~\citep{elsner2006prediction} by incorporating an elastic net penalty to address strong correlations among spatial covariates. The MA model is a climatology baseline~\citep{elsner2006prediction,klotzbach2009twenty} that predicts the hurricane count in year \( t \) as the mean over the previous five years, providing a simple yet widely used benchmark in seasonal hurricane forecasting. The UA model is a physically informed benchmark that is implemented in two steps. First, the spatial covariates SSTs are reduced to a single scalar predictor by averaging over a selected region \((60^{\circ}\text{W} - 20^{\circ}\text{W},\ 0 - 20^{\circ}\text{N})\). Second, this SST average is used alongside two scalar predictors PWS and MEI\_AMO in a Poisson regression. The selected region for computing the SST average is determined using t-test screening on the full dataset in combination with expert knowledge, which means the model is not a purely statistical predictive approach but rather an “oracle” benchmark reflecting idealized prior knowledge of the signal location. To evaluate the effect of the UA model’s ad hoc region selection, we construct a purely data-driven alternative, referred to as the T-stat method. Mirroring the UA model’s structure, we first compute location-wise t-statistics for the SST predictors using in-sample data, and designate the regions with coefficients significant at the 5\% level as active regions. The SST values over these statistically selected regions are then averaged to form a single scalar predictor, which is combined with the other covariates in the Poisson regression.

The implementations of DLC, DHS, LC, and HS differ from those in the simulation studies, as the real data setting requires a more flexible treatment of spatial variability. To ensure robust regional selection and smooth spatial variation in SST effects, we adopt a tensor product B-spline (TPB) expansion to model the spatial structure more flexibly \citep{wahba1990spline,jetter2005topics}. In the simulation study, SST was treated as a predictor on predefined regions, but in the implementation, we aim to capture finer-scale spatial variability while mitigating multicollinearity caused by strong spatial dependence. After removing missing data (primarily land regions), 1531 effective areal units remain within the study region \((100^{\circ}W - 20^{\circ}E,\ 10^{\circ}S - 60^{\circ}N)\). The spatial autoregression coefficient of SST is approximately 0.9, indicating a high degree of spatial dependence that complicates direct region selection. This strong spatial dependence increases the risk of collinearity among adjacent locations, making conventional selection methods unstable and less interpretable. To reduce spatial autocorrelation and dimensionality, we represent the SST effect using a linear combination of tensor product B-splines, which enables smooth, data-driven spatial regularization. Specifically, the SST effect at location \( j \) is expanded as:
\begin{equation}
    \label{eq::tpbsplines}
    \beta_j = \sum_{l=1}^L M_l(j)\gamma_l,
\end{equation}
where \( M_l(j) \) denotes the value of the \( l \)-th spline basis at location \( j \), and \( \gamma_l \) represents the corresponding coefficient. We construct the tensor product B-spline basis using 17 equally spaced nodes between \( 100^{\circ}W \) and \( 20^{\circ}E \), and 12 equally spaced nodes between \( 10^{\circ}S \) and \( 60^{\circ}N \), resulting in 300 cubic spline functions. By expanding \(\bm{\beta}\) in terms of spline bases, the linear predictor is given by:
\begin{equation}
\label{eq::linkwithsplines}
    \log (\bm{\theta}) = \bm{W}\bm{\alpha} + \bm{X}\bm{M}\bm{\gamma},
\end{equation}
where \( \bm{M} = [M_l(j)] \) is a \( 1531 \times 300 \) basis matrix, and \( \bm{\gamma} \) is the coefficient vector. The new design matrix \( \bm{X} \bm{M} \) has 300 columns, each representing the tensor product B-spline weighted SST over compact regions. These transformed predictors allow for flexible modeling of SST effects while ensuring smoothness and reduced multicollinearity. To further control spatial complexity and conduct region selection, we place a shrinkage prior on \( \bm{\gamma} \), which is now treated as the effect of the newly constructed spatial predictor. This formulation allows the model to adaptively learn the dominant spatial patterns of SST without imposing rigid predefined structures.

We evaluate the predictive performance of different models using four performance measures: mean absolute error (MAE), root mean squared error (RMSE), maximum absolute error (MaxAE), and standard deviation ratio (SDR). Let \( \hat{y}_t \) denote the predicted number of hurricanes in year \( t \). For the models DLC, LC, DHS, HS, and CAR, \( \hat{y}_t \) is the sample mean of the posterior predictive distribution of \( y_t \). The four measures are defined as follows:
\begin{gather*}
    MAE = \frac{1}{13}\sum_{t=2001}^{2013}|y_t - \hat y_t|, \quad
    RMSE = \sqrt{\frac{1}{13}\sum_{t=2001}^{2013}(y_t - \hat y_t)^2}, \\ 
    MaxAE = \max_{2001 \leq t \leq 2013} |y_t - \hat y_t|, \quad
    SDR = \sqrt{\frac{\sum_{t=2001}^{2013}(\hat{y}_t-\bar{\hat{y}})^2}{\sum_{t=2001}^{2013}(y_t-\bar{y})^2}}.
\end{gather*}
For first three metrics, MAE, RMSE, and MaxAE, lower values indicate better predictive performance. The standard deviation ratio (SDR) measures the relative variation in the predicted hurricane counts compared to the observed counts. If \( \text{SDR} \approx 0 \), the predicted hurricane counts are nearly constant over 2001–2013, suggesting that the model fails to capture interannual variability. If \( \text{SDR} \approx 1 \), the predicted variability closely matches that of the observed hurricane counts, indicating that the model successfully reproduces the observed variation. Note that SDR does not directly measure prediction accuracy—a model with an SDR close to 1 may still yield large prediction errors. For two models with similar prediction errors, we prefer the one with a larger SDR, as it may better capture the trend and variability of the data.

\subsection{Results}
 \begin{table}[htp!]
 \centering
 \label{tab:realdata1}
\begin{tabular}{c c c c c c c c c} 
 \hline\hline
Measure  & UA & T-stat  & \textbf{DLC} & DHS   & LC & HS    & EN  &  MA  \\ [0.5ex] 
 \hline
 MAE &  2.08 & 2.12 &  \textbf{2.13}  & 2.22 & 2.44 & 2.32  & 3.12  & 3.25 \\ 
 \hline
 RMSE & 2.43 & 2.73 & \textbf{2.57} & 2.69 & 2.89 & 2.82  & 3.84   & 3.96\\
 \hline
MaxAE & 3.71  & 5.31  & \textbf{4.12} &4.49 & 5.05 & 5.16 & 7.60  & 8.96 \\
 \hline
SDR & 0.83 & 0.61  & \textbf{0.77} & 0.71 & 0.76 & 0.73  & 0.16   & 0.28  \\
\hline
\end{tabular}
 \caption{A comparison among DLC, DHS, LC,HS, T-stat, EN, MA and the UA model.}
\end{table}

Table~\ref{tab:realdata1} summarizes the out-of-sample predictive performance of the models we considered. Although the ``oracle" benchmark model UA achieves the best performance, its advantage stems largely from leveraging expert-curated region selection and training on the full dataset—an approach that does not reflect true out-of-sample predictive performance. The slightly inferior results of the T-stat model, which selects regions using only in-sample data without domain knowledge, suggest that much of UA’s performance gain may be attributed to this retrospective use of information, rather than methodological superiority.

Among all data-driven approaches, the proposed DLC model achieves the best overall performance, with the lowest MAE and RMSE, and the highest SDR apart from UA. It also attains the second-lowest MaxAE, closely trailing the UA model. This establishes DLC as the most accurate non-oracle method for hurricane prediction. Comparing DLC with its independent counterpart LC, and DHS with HS, we observe that incorporating spatial dependence in the shrinkage prior consistently improves prediction accuracy. DHS also performs well but trails DLC slightly across all metrics, likely due to its weaker shrinkage control for small signals. Between the independent models, LC outperforms HS, mirroring the trend seen in their spatial counterparts. The elastic net (EN) performs the worst among all model-based methods, with substantially higher MAE, RMSE, and MaxAE. Its extremely low SDR (0.16) indicates under-dispersion in its predictions, effectively regressing hurricane counts to a near-constant baseline. The MA baseline also performs poorly, as expected. Overall, these results highlight the benefit of spatially dependent shrinkage in improving predictive accuracy, with DLC offering the best performance among all purely data-driven methods.

\begin{figure}[htp!]
\centering
\includegraphics[width=0.75\textwidth]{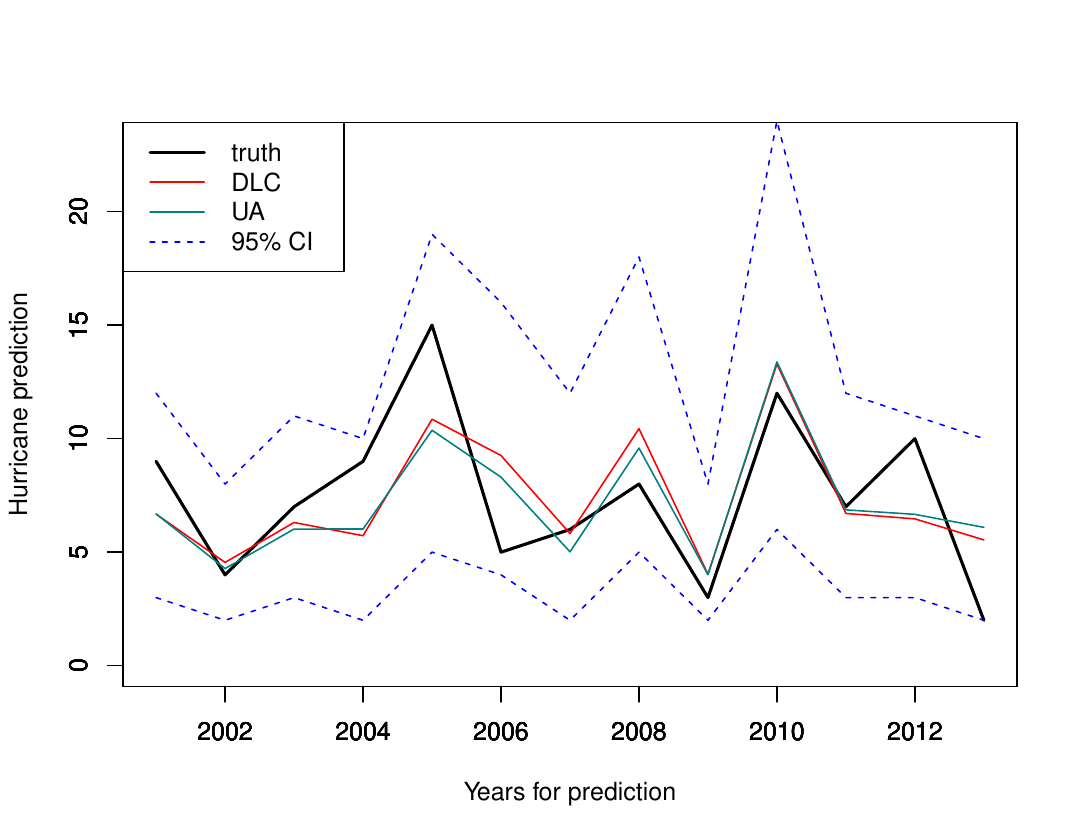}
\caption{Predicted number of hurricanes by DLC and UA.}
\label{fig::pred_DLC}
\end{figure}

Figure \ref{fig::pred_DLC} presents the predicted annual number of hurricanes from 2001 to 2013 using DLC and UA. In the figure, the black solid line represents the observed number of hurricanes (``truth"), while the red and teal solid lines denote the predicted values of DLC and UA. The blue dashed lines indicate the 95\% predictive credible interval (CI). Overall, the DLC model captures the general trend of hurricane frequency, with most predictions closely following the observed values and aligning perfectly with the ``oracle" benchmark UA. Importantly, the 95\% predictive CI successfully encompasses all observed hurricane counts, indicating that the model offers reasonable estimates of uncertainty. The largest prediction errors occur in 2005–2006, during which all models, including the UA, substantially underpredict hurricane counts. These discrepancies suggest that factors beyond SST, PWS and MEI\_AMO, such as favorable atmospheric circulation and low wind shear, might play a dominant role in those years. Such deviations highlight the challenge of predicting extreme events based solely on limited predictors. Further modeling may benefit from incorporating additional dynamic atmospheric variables \citep{bell20062004, virmani2006hurricane}.

\begin{figure}[htp!]
\centering
\includegraphics[width=0.85\textwidth]{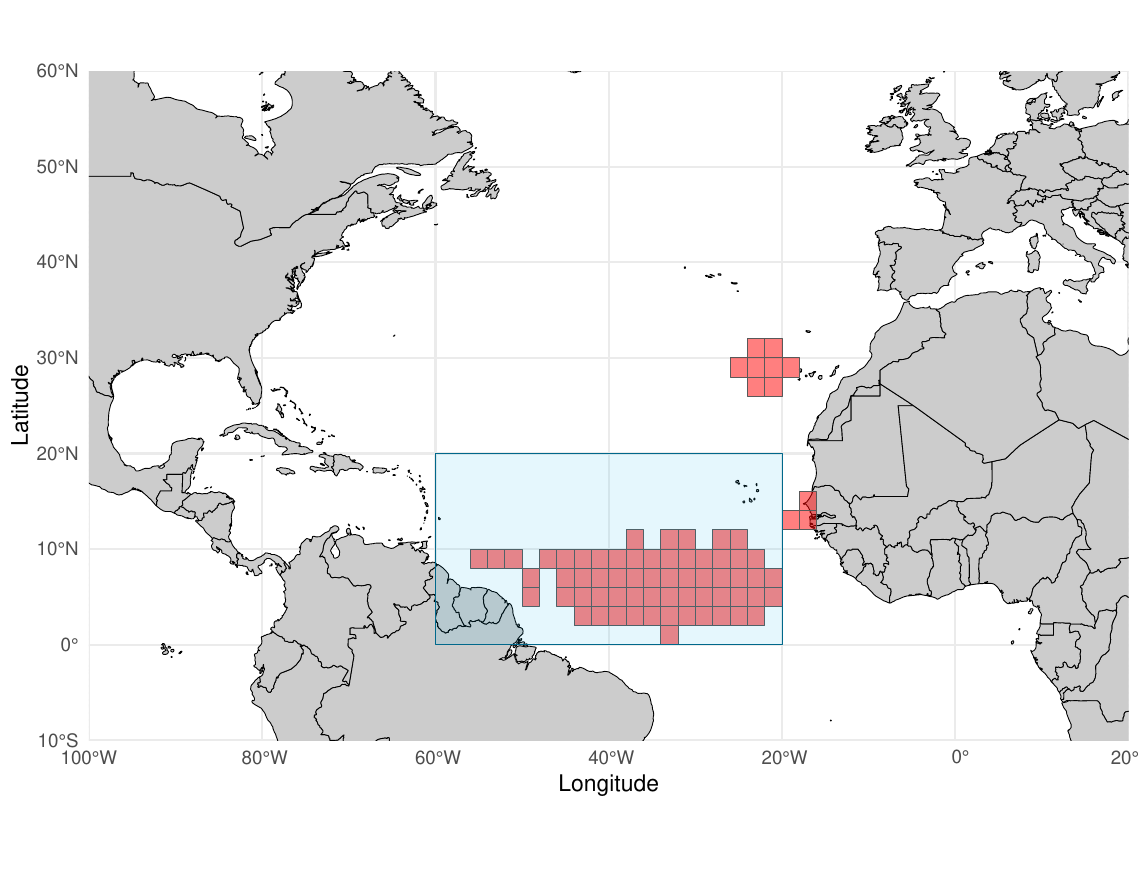}
\caption{The locations of persistently active regions in the North Atlantic selected by DLC and credible interval method (red). These regions are identified as active in at least two consecutive one-year-ahead predictions. Active regions selected by UA model are shadowed in blue.}
\label{fig::selec_DLC}
\end{figure}

Figure \ref{fig::selec_DLC} presents the regions that are persistently identified as the active regions over the years by the DLC model. The depicted regions are obtained through two steps. First, active regions are identified in each one-year-ahead prediction using credible interval method. Second, regions that remain active in at least two consecutive one-year-ahead predictions are classified as persistently active, ensuring temporal consistency and reducing the influence of short-term variability. The selected regions overlap substantially with the regions used in the UA model. However, our approach also selects a region between 30°N and 40°N near the northwest coast of Africa, which was not highlighted by the UA model. This region is located in the vicinity of the Azores Current region, suggesting distinct oceanic or atmospheric influences at these latitudes.

\begin{figure}[htp!]
\centering
\includegraphics[width=0.85\textwidth]{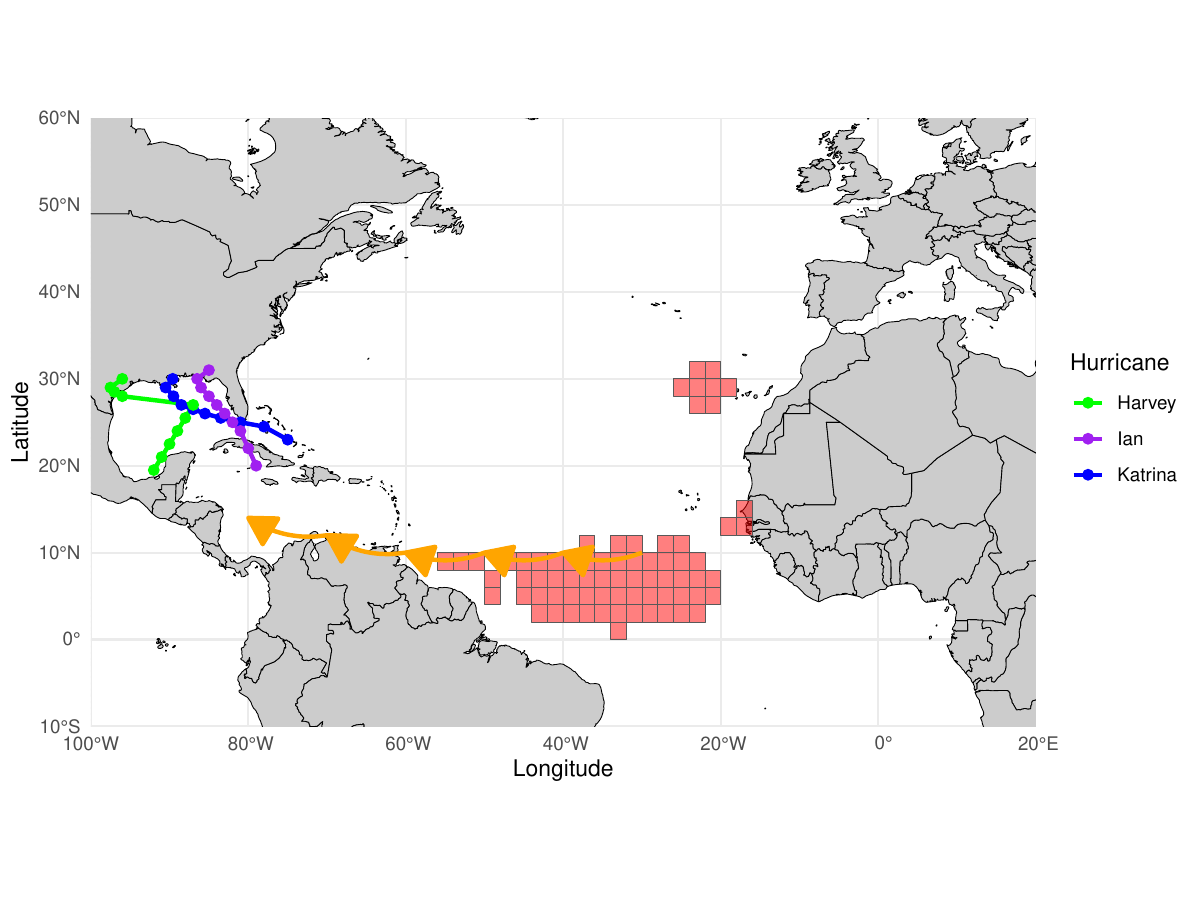}
\caption{The locations of persistently active regions (red) in the North Atlantic, along with the North Equatorial Current (NEC, orange arrows) and the tracks of three historically destructive hurricanes: Harvey, Ian, and Katrina.}
\label{fig::pathway1}
\end{figure}

Figure~\ref{fig::pathway1} shows the persistently active SST regions identified by our model, along with the tracks of three major hurricanes (Harvey, Ian, and Katrina). These regions, concentrated along the North Equatorial Current (NEC), align well with areas where warm waters are transported westward into the Caribbean. The identified active zones correspond closely to the historical development paths of these storms, suggesting that the model captures relevant heat transport patterns that support cyclone formation. Although our model is purely data-driven, the selected regions match known climatological structures that supply thermal energy for hurricane intensification \citep{Enfield1997}. The agreement between predicted active regions and observed hurricane tracks highlights the model’s capacity to recover physically meaningful spatial signals without prior expert guidance.

\begin{figure}[htp!]
\centering
\includegraphics[width=0.85\textwidth]{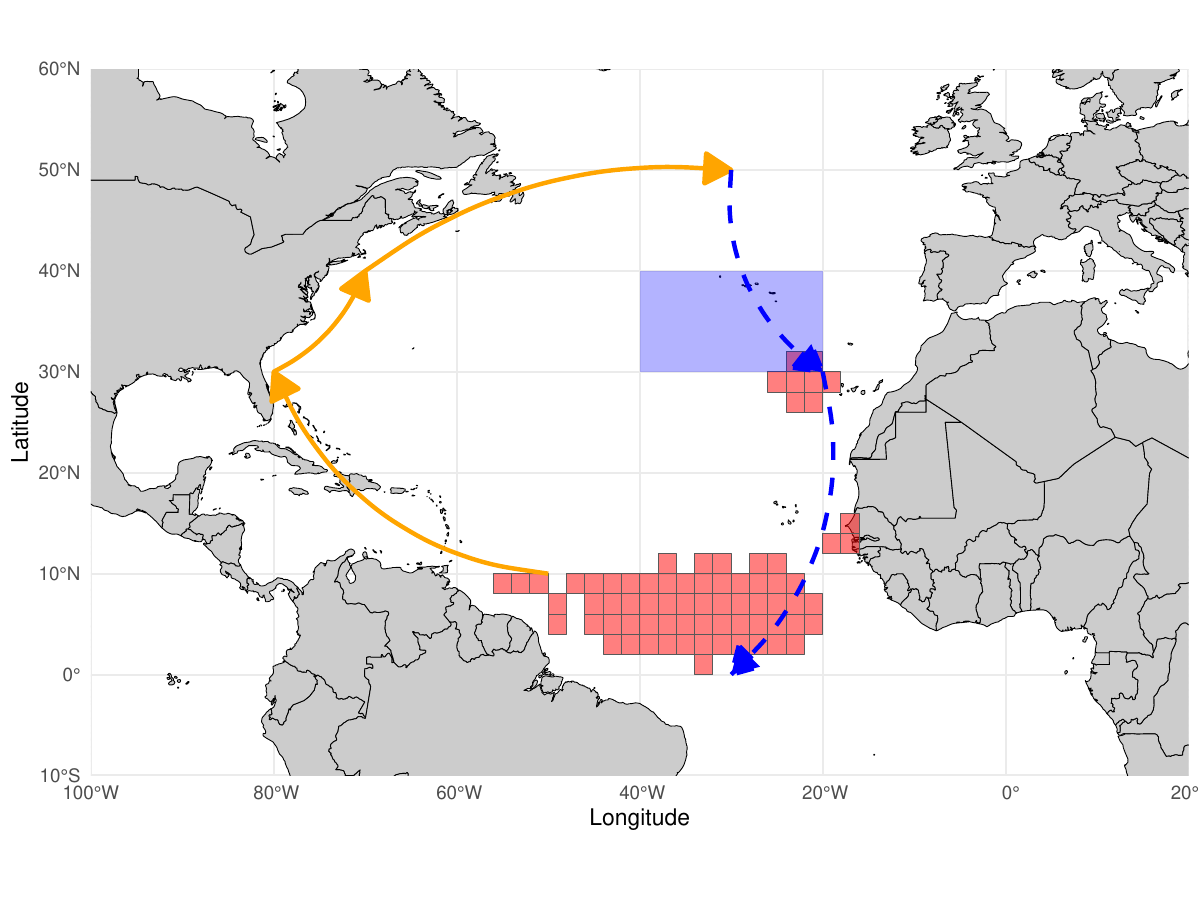}
\caption{The Atlantic Meridional Overturning Circulation (AMOC) and the Azores High in the North Atlantic region. The AMOC consists of a northward warm surface current (orange arrows) and a southward deep cold current (blue dashed arrows). The red regions are the persistently active oceanic areas identified by DLC.}
\label{fig::pathway2}
\end{figure}

Figure~\ref{fig::pathway2} further illustrates model-identified active regions near 30°N–40°N west of Africa, alongside the typical path of the Atlantic Meridional Overturning Circulation (AMOC) and the Azores High. These spatial patterns suggest potential associations between SST variability in the northeast Atlantic and downstream hurricane activity. The Azores High influences storm steering and mid-tropospheric stability, while AMOC transports oceanic heat northward. Though not explicitly modeled, these background processes align with our model’s selection of regions that may modulate hurricane frequency or intensity. Recent studies \citep{Rahmstorf2015AMOC} have raised concerns over AMOC disruptions, further motivating the importance of statistically detecting changes in these upstream areas. Our results suggest that data-driven spatial selection can capture climate-relevant structures with predictive value, offering complementary insight to physics based models.

\section{Discussions}
We propose a spatially dependent global-local (SGL) shrinkage prior for estimating the effects of spatially dependent covariates and selecting active regions in Poisson regression. This method combines global-local shrinkage priors with a CAR prior to incorporate spatial dependence. Similar to independent global-local shrinkage priors, our proposed prior effectively shrinks both signals and noises. However, the spatial extension introduces a unique capability: it adaptively shrinks the conditional covariance between noises or signals and noises, while preserving the conditional covariance between signals. This feature ensures better recovery of adjacent signals. 

Compared to independent global-local shrinkage priors, our method requires a heavier tail for the prior on the local shrinkage parameters and more mass around the origin for the global shrinkage parameter. Consequently, super heavy-tailed priors such as the log-Cauchy prior are crucial in applying the spatially dependent global-local shrinkage prior. Extensive simulation studies demonstrate the empirical performance of the proposed prior in Poisson regression. When applied to the North American hurricane count prediction problem, our model outperforms the elastic net and closely matches the performance of the benchmark "oracle." Beyond Poisson regression, the proposed SGL prior can be readily extended to other generalized linear models. Moreover, the core idea of using global-local shrinkage under structured dependence is not limited to regression settings. For example, similar global-local shrinkage principles have been applied to spatially structured small area estimation problems \citep{tang2018modeling,tang2024hierarchical}.

While our proposed SGL prior demonstrates strong performance in both estimation and region selection, DHS prior tends to produce uncompetitive selection performance. As discussed in Section~4.2, this issue stems partly from the intrinsic behavior of the horseshoe prior, which---though effective for sparse estimation---tends to over-shrink moderate signals in the presence of strong spatial dependence. In addition, the commonly used region selection strategy based on credible intervals becomes increasingly unreliable in high-dimensional settings or under strong spatial correlation. In such contexts, even small propagated signals may yield credible intervals that exclude zero, leading to false positives. An alternative selection framework that has the potential to address this is the scaled neighborhood (SN) approach~\citep{Li2010TheBE}. Under this criterion, a region $j$ is classified as inactive if the posterior probability of $\beta_j$ lying within $(- \hat{\sigma}_j, \hat{\sigma}_j)$ exceeds 0.5, where $\hat{\sigma}_j$ denotes the posterior standard deviation of $\beta_j$. This rule prioritizes concentration of posterior mass near zero over strict interval exclusion, making it more robust to small spurious signals and spatial spillover. Despite its intuitive appeal, the SN approach has received limited follow-up in the literature, with little theoretical characterization or systematic empirical evaluation beyond its original proposal.

Looking ahead, several avenues merit further investigation to enhance the applicability and performance of the proposed framework. First, while our current implementation relies on standard MCMC algorithms, the hierarchical structure of the spatially dependent global-local shrinkage prior introduces computational challenges in high-dimensional settings. Developing more efficient inference strategies---such as block-wise Gibbs sampling, Hamiltonian Monte Carlo tailored to sparse spatial priors, or variational approximations with structured dependencies---could substantially accelerate posterior sampling while preserving inferential accuracy.

Second, the current model focuses exclusively on spatially dependent covariates. In many real-world applications, especially in climate and environmental studies, the underlying signals evolve over both space and time. Extending our global-local shrinkage framework to explicitly accommodate spatio-temporal dependencies would allow for more nuanced modeling of temporally varying regional effects and improve forecasting in dynamic settings. This extension would require the construction of spatio-temporal priors that simultaneously induce sparsity, preserve signal integrity, and respect structured dependencies across both dimensions.

Third, our current framework is developed for settings in which a single spatial variable is considered as the primary high-dimensional predictor. In practice, however, multiple spatial variables may jointly influence the outcome. For example, in the hurricane prediction problem, PWS is also a spatial variable, but it is often preprocessed into a scalar summary to facilitate analysis. Incorporating multiple spatial covariates directly into the model and performing joint region selection across these fields would substantially broaden the applicability of the proposed approach. Such extension introduces new methodological challenges, including the need to model cross-variable spatial dependence, disentangle overlapping or interacting spatial signals, and design shrinkage priors that enable coordinated yet flexible selection across multiple spatial domains. Addressing these challenges would require novel prior constructions and inference strategies, and represents a promising direction for future research.

\begin{acks}[Acknowledgments]
We thank Dr.~Xubin Zeng of the Department of Hydrology and Atmospheric Sciences at the University of Arizona for kindly providing the data.
\end{acks}
\bibliographystyle{imsart-nameyear} 
\bibliography{bibliography.bib}       


\end{document}